\documentclass[aps,prl,prabib,twocolumn,showpacs,nofootinbib]{revtex4}
\usepackage{graphicx} \usepackage{amsmath} \usepackage{amssymb}
\usepackage{amsfonts} \usepackage{bm}

\begin{document}

\newcommand{\be}{\begin{equation}} \newcommand{\ee}{\end{equation}}
\newcommand{\bea}{\begin{eqnarray}}\newcommand{\eea}{\end{eqnarray}}

\title{Hydrino like states in graphene and Aharonov-Bohm field}

\author{Pulak Ranjan Giri} \email{pulakranjan.giri@saha.ac.in}

\affiliation{Theory Division, Saha Institute of Nuclear Physics,
1/AF Bidhannagar, Calcutta 700064, India}

\begin{abstract}
We study the dynamics of fermions on graphene in presence of Coulomb
impurities and Aharonov-Bohm field. Special emphasis is given to the
formation of hydrino like states and its lifting of degeneracy due
to the presence of AB field. The flux of the AB field can be tuned
to make the low angular momentum hydrino states stable against
decay. The zero limit physics of the two coupling constants $\alpha_G$ and 
$\Phi$ involved in the system is discussed.
\end{abstract}



\date{\today}

\maketitle

Graphene, a 2-dimensional form of allotrope, made up  of single layer of 
carbon atoms, has
been a platform to study the various exotic features \cite{novo2} of
fermions which is usually not seen in  most condensed matter
systems. The exotic behavior is  attributed to the unique lattice
structure of the allotrope, which can be understood from the fact
that the different layers of graphite are so loosely bound that it
slides one above another. The bonding between the atoms of the same
layer are strong enough to form a stable structure. It can be seen
when struck by a pencil lead on a paper. The mark on the paper can
be considered as a thin layer of carbon atoms. The recent
experimental formation of graphene in 2004 \cite{novo1} has made it
possible to test several  features in this system. For example, in
semiconductor or in  other material, the energy-momentum relation
for the electrons with effective mass $m^*$ are given by $E=
p^2/2m^*$. But for monolayer graphene without any external
perturbation or distortion, the dispersion relation for the
electrons near a Dirac point  are given by  $E = \pm c_G|\boldsymbol
p|$ with velocity $c_G$. This linear dispersion relation was a clue
that the charge carriers in graphene could be described by massless
Dirac equation. However, mass term for the fermions could arise if
somehow the two sub lattices of the graphene are made
distinguishable.

One of the exotic behaviors is that, the mobility of charge carriers
in graphene is approximately independent of the carrier density. The
constancy of the mobility means  that the conductivity in graphene
is proportional to the carrier density. The list of unusual
phenomena in graphene is even extended in quantum Hall effect, which
can not be explained by non-relativistic quantum mechanics. The
integer quantum Hall effect occur at half integral values of the
filling factor, which is therefore termed as anomalous
\cite{novo2,li}. This effect is common in massless monolayer
graphene and massive bilayer graphene.  The Landau levels which are
important to describe quantum Hall effect are also different from
what is seen in non-relativistic quantum mechanics. In constant
magnetic field $B$ background perpendicular to the plane, the
electron motion in metal or semiconductor  is quantized as  $E=
\hbar\omega^*(n+1/2), n=0, 1, 2,..$. The holes also have  same
behavior but with negative energy. Here $\hbar$ is the reduced Plank
constant and $\omega^*= eB/m^*$ is the cyclotron frequency. But
graphene has completely different Landau levels \cite{li}. In
monolayer graphene, it is given by
$E_G=\mbox{sgn}(n)c_G\sqrt{2e\hbar|n|B}$. The occurrence  of the
$E_G=0$ states are the consequence of chirality \cite{jackiw1} of the Dirac
fermions. The degeneracy of the
 $E_G=0$ state is however same as the $E_G \neq 0$ state.
On the other hand, for
 bilayer graphene  the spectrum is of the form $E_G= \mbox{sgn}(n)\hbar
\omega_G\sqrt{|n|(|n|+1)}, n= 0, \pm 1, \pm 2,...$. The $E_G=0$
levels are still present but this time the degeneracy  is double
than that of $E_G\neq 0$ levels.

All the above features are indicative of the fact that the dynamics
of charge carriers in graphene is  described by Dirac fermions. The
tight binding model \cite{gordon} of graphene indeed show that the
dynamics of electron and hole are dictated by (2+1)-dimensional
Dirac fermion in the continuum limit of the lattice spacing.
Graphene  is thus considered to be a condense matter analog of QED
(see the introduction of \cite{levitov}). However, it has lot of
differences with the standard QED. Depending upon the situation,
fermions could have mass or zero rest mass as mentioned before. The
velocity of the zero rest mass fermions are $c_G\approx c/300$,
which is much less than the light velocity $c$ in vacuum. Due to
this small Fermi velocity $c_G$, the fine structure constant is much
higher $\alpha_G\approx 300\alpha$. Obviously the perturbation
scheme of QED seems to be not applicable to graphene system due such
a  high coupling constant $\alpha_G$.

In this work, we report another possibility of exotic phenomena of
massive fermions on graphene, which is the formation of  hydrino
like states. But due to the high coupling constant $\alpha_G$, the
so called hydrino states may have complex eigenvalues for low
angular momentum  values, which will make it dissipative. We
therefore consider a Aharonov-Bohm (AB) field, which will rescue the
low angular momentum states and will form stable bound states.
Hydrino state is the tightly bound states of Hydrogen atom with
binding energy $E_B
> 13.6 eV$, where normal ground state energy for Hydrogen atom is
$E_G= -13.6 eV$. Experimental study has also been reported recently
in support of this hydrino states (see \cite{rathke} and its refs.).
Theoretical model of hydrino states can be found by solving
Klein-Gordon equation in 3 space dimensions. Recently  it has been
shown that, in $(2+1)$-dimensions, Dirac equation may give rise to
hydrino like states \cite{dombey}. Besides the hydrino states, the
large fine structure constant $\alpha_G$ of graphene  also may make
the normal bound state eigenvalue complex. Our inclusion of a AB
field will also make the normal bound states stable.

The existence of deeply bound states can be explained in terms of a
consistent mathematical formulation known as self-adjoint extensions
\cite{reed}. In this method the commonly used Dirichlet boundary
condition at origin of the wave-function, that  the wave-function is
zero at origin, is generically extended to a less restricted
boundary condition. Usually  the wave-function at the origin is
considered to have a certain power law behavior without  violating
the square-integrability condition. This larger domain then allows
other states to immerse out as a viable solutions.

We start our discussion with the fermions on graphene in presence of
a gauge potential $A_\mu$, described by a two component Dirac
equation
\begin{eqnarray}
\left(i \displaystyle{\not}\partial +  \displaystyle{\not}A -M
\right)\psi(t;r,\phi)=0 \label{d1}
\end{eqnarray}
We set the unit in which Fermi velocity $c_G\approx 10^6m/s =1$ and
$\hbar=1$. Here
$\displaystyle{\not}\partial=\gamma^\mu\partial_\mu$,
$\displaystyle{\not}A= \gamma^\mu A_\mu$ and we take the
representation of the gamma matrices in terms of Pauli matrices as
$\gamma^0= \sigma^3$, $\gamma^1= i\sigma^1$ and $\gamma^2=
i\sigma^2$. Since we are considering  Coulomb impurities and
Aharonov-Bohm field, our gauge potential $A_\mu$ is  the Coulomb
potential $A_0= \alpha_G/r$ and AB field $A_\phi= -\Phi/r$. The mass
$M$ of the fermion could generate due to the external perturbation,
which makes the two sub-lattices of the honeycomb lattice structure
of graphene distinguishable or by other means \cite{balatsky}. 
The the hopping of the electrons
between two sub-lattices makes the energy difference nonzero.  In
order to solve (\ref{d1}) we need to know the condition under which
the radial hamiltonian extracted from (\ref{d1}) is symmetric.  This
condition can be easily found to be
\begin{eqnarray}
\lim_{r\to 0} r\phi_1^\dagger(r)i\sigma^2\phi_2(r)=0\,,
\label{condl}
\end{eqnarray}
where $\phi_1$ and $\phi_2$ are two component radial spinors. To
construct a domain for our problem we need to construct it in such a
way that it respect the constraint (\ref{condl}) and at the same
time it becomes equal to its adjoint domain. We mentioned in the
introduction that, the mathematical formulation which serve this
purpose is known as self-adjoint extensions and it has been recently
discussed in case of graphene with Coulomb impurities in
\cite{kumar2} and a generalized boundary condition for the bound
state solution  has been obtained.

We do not go into detail analysis of it, rather we give here an
intuitive argument to show that the boundary condition can be
parameterized by $U(1)$ group. The short distance behavior of the
two independent  radial wave-functions $\psi_1(r)$ and $\psi_2(r)$
has the form
\begin{eqnarray}
\nonumber \lim_{r\to 0}\psi_1(r) &\sim& \mathcal{A} r^{-1/2 + \xi}\\
 \lim_{r\to 0}\psi_2(r) &\sim& \mathcal{B} r^{-1/2 - \xi}\,,
\label{shortdistancel}
\end{eqnarray}
where $\mathcal{A}$ and $\mathcal{B}$  are the two component
constants, $\xi= \sqrt{(m+\Phi)^2-\alpha_G^2}$, and $m= \pm 1/2, \pm
3/2,...$. For $\Phi=0$,  (\ref{shortdistancel}) is reduced to the
short distance behavior of radial function of \cite{dombey}.
 Note that both the solutions are
square-integrable at $r=0$ if
\begin{eqnarray}
0\leq  2|Re(\xi)| <1
\label{squareintegrablel}
\end{eqnarray}
Technically, the point $r\to 0$ is called  quantum mechanically
incomplete \cite{reed} and the potential corresponding to the radial
Hamiltonian is called in limit circle case. On the other hand  only
one solution is square-integrable near $r \to\infty$. So the
potential is in limit point case at infinity. This means, there can
have 1-parameter family of self-adjoint extensions. Since both
solutions are square-integrable at $r\to 0$, then any linear
combinations of the two will also be square-integrable. We take a
generalized boundary condition of the form
\begin{eqnarray}
\lim_{r\to 0} \psi(r)\equiv  \mathcal{C}_1(\theta) r^{-1/2 + \xi} +
\mathcal{C}_2(\theta) r^{-1/2 -\xi}\,,
\label{generalcondl}
\end{eqnarray}
where the constants, $\mathcal{C}_1(\theta)$ and
$\mathcal{C}_2(\theta)$, which are now dependent on the free
parameter $\theta$ of the $U(1)$ group, be such that
(\ref{generalcondl}) satisfies the constraint (\ref{condl}). Then
$\theta$ is called the self-adjoint extension parameter. The normal
bound state solutions can be obtained by setting the coefficient of
the singular term of (\ref{generalcondl}) to zero, i.e.,
$\mathcal{C}_2(\theta)=0$. This gives the bound state solution of
the form
\begin{eqnarray}
E_{n,m}= M \mbox{sgn}
(\alpha_G)/\sqrt{1+\alpha_G^2/(n+\sqrt{(m+\Phi)^2-\alpha_G^2})^2}.
\label{bsl}
\end{eqnarray}
Note that for $\Phi=0$ the above solution reduces to the result
\cite{novikov,dombey}. Now comes the exotic solutions of the
problem, which can be obtained by setting the 1st term of
(\ref{generalcondl}) to zero but keeping the singular term. In this
case the solutions are square-integrable and the bound state
eigen-value is given by
\begin{eqnarray}
E_{n,m}= M \mbox{sgn}
(\alpha_G)/\sqrt{1+\alpha_G^2/(n-\sqrt{(m+\Phi)^2-\alpha_G^2})^2}\,,
\label{bs2}
\end{eqnarray}
which reduces to the result of \cite{dombey} for $\Phi=0$. This is
the so called hydrino like states. Note that in presence of the
magnetic flux, the angular momentum degeneracy is lifted.  As is
evident the states with spectrum (\ref{bs2}) are tightly bound than
the states with eigenvalues (\ref{bsl}). we have two parameters in
our system, one is fine structure constant $\alpha_G$ and other one
is the flux $\Phi$, whose limiting behavior can be investigated. We
see that the limit $\Phi \to 0^\pm$ is smooth and from (\ref{bsl})
and (\ref{bs2}) we get
\begin{eqnarray}
\lim_{\Phi\to 0^+}E_{n,m} = \lim_{\Phi\to 0^-}E_{n,m}=
\lim_{\Phi=0}E_{n,m}\label{bslimit}
\end{eqnarray}
But the similar limit for $\alpha_G$ shows that
\begin{eqnarray}
\lim_{\alpha_G\to 0^+}E_{n,m} \neq \lim_{\alpha_G\to 0^-}E_{n,m}\neq
\lim_{\alpha_G=0}E_{n,m}\label{bslimit1}
\end{eqnarray}
Note that the first inequality of (\ref{bslimit1}) can be obtained
from (\ref{bsl}) or (\ref{bs2}), but to get the second inequality we
need to compare our result with \cite{gerbert}.

To summarize, we considered the quantum dynamics of fermion on
graphene in presence of Coulomb impurities and AB field background.
Possible generalized boundary condition has been established based
on intuited arguments. The fermions are trapped  in the combined
potential. We showed that the deepest bound state solutions, known
as hydrino like states, are nothing but a  consequence of a specific
choice of boundary condition. It can be rigorously shown by
self-adjoint extensions. The limit $\Phi \to 0$ is shown to be
smooth but the limit of the fine structure constant is not smooth so
can not be used to get the dynamics of the fermions in only AB field
background. Due to large value of the  fine structure constant
$\alpha_G$ of graphene, the low angular momentum states $E_{n,m}$
may become complex when only Coulomb field is present. But the
presence of the AB field may rescue it and make the states stable.


\begin{thebibliography}{99}
\bibitem{novo2} K. S. Novoselov et al, Nature {\bf 438}, 197 (2005).

\bibitem{novo1} K. S. Novoselov et al, Science {\bf 306}, 666 (2004).

\bibitem{li} G. Li and E. Y. Andrei, Nature {\bf 3}, 623 (2007).

\bibitem{jackiw1} R. Jackiw and  S.-Y. Pi, Phys. Rev. Lett. {\bf 98}, 
266402 (2007). 


\bibitem{gordon} G. W. Semenoff, Phys. Rev. Lett. {\bf 53}, 2449
(1984).

\bibitem{levitov} A. V. Shytov, M. I. Katsnelson and L. S. Levitov,
Phys. Rev. Lett. {99}, 236801 (2007).

\bibitem{rathke} A Rathke, New J. Phys. {\bf 7}, 127 (2005).



\bibitem{dombey} N. Dombey, Phys. Lett. A {\bf 360}, 62 (2006).

\bibitem{reed} M. Reed and B. Simon,  {\it Fourier Analysis, Self-Adjointness}
  ( New York :Academic, 1975 ).

\bibitem{balatsky} T. O. Wehling, I. Grigorenko, A. I. Lichtenstein and 
A. V. Balatsky, arXiv:0803.4427v1 [cond-mat.mes-hall].


\bibitem{kumar2} K. S. Gupta, S. Sen, arXiv:0808.2864v1 [hep-th].

\bibitem{gerbert} Ph. de S. Gerbert, Phys. Rev. {\bf D40}, 1346 (1989).

\bibitem{novikov} D. S. Novikov,  Phys. Rev. {\bf B76}, 245435 (2007).




\end{thebibliography}
\end{document}